\documentclass{article}
\usepackage{amsmath}

\input{tcilatex}

\begin{document}

\title{\textbf{Specificity of Phase Transition of Quasi-Spin System in
Two-Quantum Exchange with Thermostat}}
\author{Nicolae Enaki$^{\text{a}}$\ and Vitalie Eremeev$^{\text{a,b}}$ \\
\\
$^{\text{a}}$Institute of Applied Physics,\\
Academy of Sciences of Moldova, \\
str. Academiei 5, Chisinau MD-2028, Republic of Moldova\\
Fax: (373 22) 739805\\
E-mail: enakinicolae@yahoo.com\\
\\
$^{\text{b}}$Faculty of Informatics Sciences and Engineering,\\
Free International University of Moldova,\\
str. Vlaicu Parcalab 52, Chisinau MD-2012, Republic of Moldova}
\maketitle

\begin{abstract}
The quantum phase transition of the system of \textit{N} radiators in case
of two-photon exchange interaction with cavity electromagnetic field is
considered. It is shown that for this system the atom-field exchange
integral increases with the increase of temperature. This effect generates
peculiarities in phase transition picture. The process of nonlinear exchange
between the quasi-spins leads to the increase of critical temperature as
compared to traditional second order phase transition of quasi-spin system.

\textbf{Keywords:} Phase transition; Two-quantum interaction; Critical
temperature; Quasi-spin system

\textbf{PACS:} 05.70.Fh; 05.30.-d
\end{abstract}

The cooperative emission phenomenon for dipole-forbidden transitions of
inverted system of radiators can be observed in the processes of two-photon
spontaneous emission in free space [1] or cavity [2]. It is important from
the physical point of view to study the cooperative phenomena in a larger
aspect of statistical physics. A great interest in the investigation of
statistical and thermodynamical properties of quantum systems represents the
existence of phase transition such as the order-disorder transition [3]. For
example a radiative system like Dicke model [4] exhibits phase transition
from the phase where the radiators emit non-correlated to the phase where
cooperativity between the radiators is established in such a way that the
intensity of emission becomes proportional to the square number of
radiators, named as super-radiance. In recent years, the study of the
phenomena which appear in physical systems when the electronic or atomic
subsystems are in non-linear interaction with the thermostat (large
subsystem) is of particular interest. For example, the non-linear mechanism
of superconductivity [5, 6], bipolaron effects [7], two-photon emission
[1,2,8] and other non-linear phenomena. Very recently, authors in [6]
obtained some interesting results for the feature of phase transition for
superconductivity in the case when a non-linear exchange interaction of the
electronic subsystem with the phonon subsystem exists. Therefore this paper
is devoted to the non-traditional behavior of order parameter for two-phonon
superconductivity thus stating that the enhancing of the critical
temperature becomes possible in comparison with usual BCS model. So we
concluded that the existence of dependence of two-phonon exchange integral
on temperature by means of the average number of phonons changes essentially
the feature of phase transition and consequently the value of critical
temperature with possibility to enhance this. The peculiarities of
two-photon exchanges between two subsystems was discussed in papers [8, 9].
In paper [9] it was shown that in special conditions the two-photon quantum
oscillator can turn from incoherent to coherent stimulated emission near the
threshold in the same fashion as the one-photon laser. Thus were obtained
special ignition and evolution processes (similar to phase transition, see
[10]) of two-photon laser from the incoherent to stable coherent emission in
the case when the two-photon losses in multi-mode cavity prevail over the
traditional one-photon losses in the resonator mirrors.

In this paper we are interested to study the cooperative phase-transition of
the system of \textit{N} radiators that interact with the cavity thermostat
via the two-quantum exchange processes. The two-photon process is possible
when the one-photon interaction with the thermostat is forbidden. Such a
possibility can be realized in microcavities with Rydberg atoms similarly to
the experiments with two-photon maser effect described in [11]. For a large
number of atoms in the cavity the phase transition through the two-photon
interaction with the cavity modes can be possible. In this case the
anomalous behavior of temperature dependence of the super-radiant order
parameter (e.g. atomic polarization) is investigated. As the two-photon
exchange integral between the radiators strongly depends on the temperature
via the average number of quanta similar to the two-phonon superconductivity
[6], so that the increasing of the order parameter with temperature in
quasi-spin systems is observed. This effect leads to the increase of value
of the critical temperature as compared to the traditional phase transition
of quasi-spin system in which the exchange integral between the spins does
not depend on the temperature.

Let us consider an ensemble of radiators in the micro-cavity that enters
into two-photon resonance with cavity electromagnetic field. In bad cavity
limits when the two-photon spontaneous emission rate is less than the losses
of cavity we can adiabatically eliminate the EMF operators and obtain the
master-equation for the atomic system discussed in [12].

\begin{eqnarray}
\dot{\rho} &=&i\left( \omega _{21}-2\bar{n}^{2}\chi \right) \left[ \rho
,R_{z}\right] -i\left( 1+2\bar{n}\right) \chi \left[ \rho ,R^{+}R^{-}\right]
\notag \\
&&-\left( 1+\bar{n}\right) ^{2}\gamma \left( \rho R^{+}R^{-}-2R^{-}\rho
R^{+}+R^{+}R^{-}\rho \right)  \notag \\
&&-\bar{n}^{2}\gamma \left( \rho R^{-}R^{+}-2R^{+}\rho R^{-}+R^{-}R^{+}\rho
\right) .  \TCItag{1}
\end{eqnarray}%
Here the following parameters are defined: $\omega _{21}$is the atom
transition frequency, $\bar{n}$ is the average number of emitted photons in
the cavity. The parameter $\chi =\frac{A(k)}{\hbar ^{4}}\frac{2\omega
_{k}-\omega _{21}}{(\omega _{21}-2\omega _{k})^{2}+4\Gamma ^{2}}$ describes
the cooperative atom-atom interaction through the cavity EMF vacuum and $%
\gamma =\frac{A(k)}{\hbar ^{4}}\frac{2\Gamma }{(\omega _{21}-2\omega
_{k})^{2}+4\Gamma ^{2}}$ is a spontaneous emission rate of single atom in
the cavity, where $A(k)$ is considered for the scheme of transitions taken
into account in [12] in the form of

\begin{equation*}
A(k)=\left[ \sum_{\alpha }\frac{\left( \mathbf{g}_{k}\mathbf{d}_{\alpha
1}\right) \left( \mathbf{g}_{k}\mathbf{d}_{\alpha 2}\right) \left( \omega
_{\alpha 1}-\omega _{2\alpha }\right) }{\left( \omega _{2\alpha }-\omega
_{k}\right) \left( \omega _{\alpha 1}-\omega _{k}\right) }\right] ^{2}\text{.%
}
\end{equation*}%
Here $\mathbf{d}_{\nu \beta }$ is the moment of dipolar transition between
the virtual states $|\alpha \rangle $ and states $|\beta \rangle $ ($\beta
=1,2$); $\mathbf{g}_{k}=\sqrt{2\pi \hbar \omega _{k}/V}\mathbf{e}_{\lambda }$
is atom-cavity interaction constant, where $\mathbf{e}_{\lambda }$ is
polarization and $V$ is volume of micro-cavity. In Eq.(1) are considered the
following cooperative quasi-spin operators $R^{+}$ , $R^{-}$ and $R_{z}$ are
defined like in [13] which are obeying the commutation relations: $\left[
R^{+},R^{-}\right] =2R_{z}$ and $\left[ R_{z},R^{\pm }\right] =\pm R^{\pm }$.

From the above master equation it is observed that for a larger value of the
detuning $\delta =2\omega _{k}-\omega _{21}$ comparatively to the cavity
damping rate\ $\Gamma $, the value of two-photon decay rate, $\gamma $, is
less than the absolute value of exchange integral, $\chi $. In this case $%
\chi /\gamma $ $\gg 1$ and we resume that the Hamiltonian part which
describes the two-photon exchange interaction between the radiators
predominates the losses part in the master equation (1). The effective
Hamiltonian that describes this interaction can be expressed as

\begin{equation}
H^{\mathrm{eff}}=\omega R_{z}-\lambda R^{+}R^{-}\text{,}  \tag{2}
\end{equation}%
where the following notations are introduced : $\omega =\hbar \left( \omega
_{21}-2\bar{n}^{2}\chi \right) $ and $\lambda =\hbar \chi \left( 1+2\bar{n}%
\right) $. Here $\lambda $ represents the energetic exchange integral
between the radiators and it is observed that this parameter depends on the
average number of emitted photons in the cavity. We emphasize that in
comparison with phase transition of traditional quasi-spin system like Dicke
model in our case the exchange integral depends on temperature through the
number of photons, $\bar{n}=\left( \exp \left[ \hbar \omega _{k}/\theta %
\right] -1\right) ^{-1}$, $\theta =\kappa T$, where $\kappa $ is Boltzmann
constant. Using the conservation of Bloch-vector at low temperature $%
R^{2}=R_{z}^{2}-R_{z}+R^{+}R^{-}$ (where $R$ is a constant), we can express
the Hamiltonian (2) through the operator $R_{z}$

\begin{equation}
H^{\mathrm{eff}}=\varpi R_{z}-\lambda \left( R+R_{z}\right) \left(
R-R_{z}\right) \text{,}  \tag{3}
\end{equation}%
where $\varpi =\omega -\lambda =\hbar \left[ \omega _{21}-\chi -2\chi \bar{n}%
\left( 1+\bar{n}\right) \right] $. Thus from the condition of existence of
the minimum value of the Hamiltonian on the operator $R_{z}$, the value $%
\tilde{R}_{z}$ can be obtained in which the system relaxes

\begin{equation}
\tilde{R}_{z}=-\frac{\varpi }{2\lambda }  \tag{4}
\end{equation}%
and therefore the value of parameter $\lambda $ must be positive from the
condition of existence of absolute minimum value of the Hamiltonian. This is
possible when the detuning $\delta $ takes positive values. As the ratio $%
\varpi /\lambda $ takes arbitrary values the restriction of values for
parameter $\tilde{R}_{z}$ can be obtained taking into account the condition
of conservation of Bloch-vector, where $R_{\max }^{2}=j(j+1)$. From this
expression follows that $R_{z}$ must take the values between $-j$ and $j$,
i.e. $-j<\left\langle R_{z}\right\rangle <j$ . Here $j=N/2$ represents the
maximum value of Bloch-vector, $N$ is the number of atoms in the cavity.
Thus, for the case of a large number of radiators from the inequality of $%
R_{z}$ and Exp. (4) it follows that the two-photon exchange integral must
satisfy the inequality $-2j\lambda <\varpi <2j\lambda $. The fact that $%
\lambda >0$ is very important, because in this case the last term in the
effective Hamiltonian (2) will imply the tendency to the order phase of the
system of radiators. Thus, for the case when $\lambda <0$ the phase
transition is not possible.

Let us consider that $R^{+}R^{-}=\sum \sum_{k,l}\sigma _{k}^{+}\sigma
_{l}^{-}=(R_{z}+j)+\sum \sum_{k\neq l}\sigma _{k}^{+}\sigma _{l}^{-}$, where
$\sigma _{k}^{\pm }$ are the Pauli spin operators. Here the first term
describes the interaction of quasi-spins with self radiation field and is
similar to traditional Lamb shift. The last term describes the interaction
of the $k$ quasi-spin with the field of other $l$ quasi-spins, while $k\neq
l $. For large value of $j$ this term can be approximated in the
thermodynamic limit (when $\lim \left( N/V\right) =\mathrm{const}$, while
cavity volume is greater than the volume per atom in the cavity) in the form
of

\begin{equation}
\dsum\limits_{l=1}^{N}\sum_{k=1,k\neq l}^{N}\sigma _{k}^{+}\sigma _{l}^{-}=%
\sqrt{N}\dsum\limits_{m=1}^{N}\left( C\sigma _{m}^{+}+C^{\ast }\sigma
_{m}^{-}\right) -NCC^{\ast }\text{,}  \tag{5}
\end{equation}%
so that the quadratic Hamiltonian (2) can be approximated in the manner of
Bogolyubov method [14] by a linear model Hamiltonian.

In the mean field approximation theory the atoms in the system can be
regarded as identical radiators so that the spin operators $\sigma _{m}^{\pm
}$ \ do not depend on the atomic label. Thus the Hamiltonian (2) according
to expression (5) takes the following form
\begin{equation}
H_{A}=N\varpi J_{z}-N\lambda \left( CJ^{+}+C^{\ast }J^{-}\right) +N\lambda
CC^{\ast }\text{,}  \tag{6}
\end{equation}%
where the number-operator $C\equiv \left\langle J^{-}\right\rangle $ plays
the role of order parameter of the system of radiators; here new operators $%
J^{\pm }=R^{\pm }/\sqrt{N}$ and $J_{z}=R_{z}/N$ are defined which obey the
commutation relations similar to the operators for radiators.

Further we would like to describe the feature of phase transition of
quasi-spin system from uncorrelated to correlated state where the
cooperativity between the radiators is established as temperature
dependence. The expression for the order parameter, $C$, can be found from
the condition of absolute minimum of free energy since we take into account
the thermodynamic analysis of the system. As was shown in the past in many
papers related to the subject of phase transition for Dicke model [4] the
free energy for the system described by Hamiltonian (2) coincides in the
thermodynamic limit with the free energy calculated by model Hamiltonian (6)
choosing the values of parameters $C$ and $C^{\ast }$ from the condition of
absolute minimum of free energy. Denoting the point of minimum by $\tilde{C}$
we have $\underset{N\rightarrow \infty }{\lim }F_{N}[H^{\mathrm{eff}%
}]=F_{\infty }[H_{A}(\tilde{C})]$ or in other words

\begin{eqnarray}
F[H_{A}(C)] &=&-\theta \log \left( \mathrm{Tr}\exp \left[ -H_{A}/\theta %
\right] \right)  \notag \\
&=&-N\theta \log \left( 2\cosh \frac{\sqrt{\varpi ^{2}+4\lambda
^{2}\left\vert C\right\vert ^{2}}}{2\theta }\right) +\lambda N\left\vert
C\right\vert ^{2}\text{.}  \TCItag{7}
\end{eqnarray}

Thus on condition that $\left( \partial F/\partial C\right) _{C=\tilde{C}}=0$%
, the value of order parameter, $\tilde{C}$, can be found by solving the
following transcendental equation

\begin{equation}
\left\vert \tilde{C}\right\vert =\frac{\lambda \left\vert \tilde{C}%
\right\vert }{\sqrt{\varpi ^{2}+4\lambda ^{2}\left\vert \tilde{C}\right\vert
^{2}}}\tanh \frac{\sqrt{\varpi ^{2}+4\lambda ^{2}\left\vert \tilde{C}%
\right\vert ^{2}}}{2\theta }  \tag{8}
\end{equation}

The critical temperature is calculated from the consideration that for $%
\theta =\theta _{\mathrm{cr}}$, the order parameter $\left\vert \tilde{C}%
\right\vert =0$ and the value of this temperature is found by solving the
equation

\begin{equation}
\tanh \frac{\varpi _{\mathrm{cr}}}{2\theta _{\mathrm{cr}}}=\frac{\varpi _{%
\mathrm{cr}}}{\lambda _{\mathrm{cr}}}\text{,}  \tag{9}
\end{equation}%
where $\varpi $ and \ $\lambda $ from the above definitions are functions of
temperature and thus will depend on the critical temperature. Hence we see
that from Eq. (9) it is not possible to find the explicit expression for
critical temperature but the equation can be solved numerically. As the
function hyperbolic tangent takes the values between $-1$ and $1$ one can
find the relations between $\chi $ and $\omega _{21}$ which satisfy $2\bar{n}%
_{\mathrm{cr}}^{2}<\omega _{21}/\chi <2(1+\bar{n}_{\mathrm{cr}})^{2}$ and
thus the critical temperature is in strong connection with values of
parameters $\chi $ and $\omega _{21}$.

Exp.(8) clearly defines the important role of the two-photon exchange
integral, $\lambda $, under the temperature dependence of order parameter, $%
C $. The intrinsic temperature dependence of the physical quantities $%
\lambda $ and $\varpi $ strongly influences the behavior of phase transition
and new features are manifested in comparison with traditional phase
transition in quasi-spin systems when exchange integral is constant. This
peculiarity is observed from Fig.(1) in which we compare the traditional
behavior of second order phase transition with temperature dependence of the
order parameter discussed in our model, see Exp.(8).\textrm{\ }It\textrm{\ }%
is an interesting fact that, in the case of our model the temperature
evolution of the order parameter in dependence with the value of the
parameter $\chi /\omega _{21}$ differs substantially from the evolution of
the order parameter calculated for traditional quasi-spin model which is
obtained from Exp.(8) omitting the term which is proportional to the mean
number of photons, $\bar{n}$. Let us consider, for example, the case when $%
\chi /\omega _{21}=0.5$ for which the phase transition in the usual
quasi-spin model is absent. As follows from the numerical results of Eq.(8),
the phase transition is possible in this case, plotted in Fig. (1a). From
this figure we observe that the order parameter appears from vacuum
fluctuations of cavity field which increases with the increase of the
temperature. Achieving the maximum value the order parameter decreases to
zero value in critical point computed from Exp.(9). The small enhancing of
value of parameter, $0.5<\chi /\omega _{21}<0.6$,\ leads to the appearance
of phase transition in traditional quasi-spin model which exhibits a
critical temperature to a lesser value in comparison with phase transition
in case of proposed model (see Fig.1b, c). Further enhancing of value of $%
\chi /\omega _{21}$ will evidence inessential differences between the
compared models and is of no interest for numerical and graphical
interpretation.

Another interesting result refers to the mean value of atomic population, $%
\langle R_{z}\rangle $, which can be calculated by the standard definition, $%
\langle R_{z}\rangle =\mathrm{Tr}\left\{ R_{z}\rho \right\} $, or using the
expression of free energy the following relation $\partial F[H_{A}]/\partial
\tilde{\omega}=-N\mathrm{Tr}\left\{ \ldots J_{z}\right\} /\mathrm{Tr}\left\{
\ldots \right\} =N\langle J_{z}\rangle =\langle R_{z}\rangle $ is observed,
consequently obtaining%
\begin{equation}
\langle R_{z}\rangle =-\frac{1}{2}\frac{\varpi }{\sqrt{\varpi ^{2}+4\lambda
^{2}\left\vert \tilde{C}\right\vert ^{2}}}\tanh \frac{\sqrt{\varpi
^{2}+4\lambda ^{2}\left\vert \tilde{C}\right\vert ^{2}}}{2\theta }  \tag{10}
\end{equation}

This value can be compared with the value calculated by Exp.(4) in order to
appreciate the differences between the results obtained using the exact
effective Hamiltonian (2) and the asymptotic Hamiltonian (6).

From Fig.2 we see that values of atomic population calculated in two
different manners coincide in the domain of temperatures close to the
critical temperature. This result confirms the accuracy of above
calculations where the asymptotic Hamiltonian (6) was considered.

In this paper we discussed the contribution of two-photon exchange mechanism
between the quasi-spins to the behavior of phase transition of such system.
Therefore obtaining an intrinsic temperature dependence of two-photon
exchange integral, the radiator system will exhibit a different fashion of
phase transition in comparison with traditional second order phase
transition of spin systems. The anomalous temperature dependence of order
parameter in the case of proposed model is influenced by the appearance in
the expressions of energy, $\omega $, and exchange integral, $\lambda $, of
additional terms which are proportional to the mean number of photons. From
the numerical simulation we obtained that the critical temperature can be
enhanced taking into account the non-linear exchange process between the
radiators in cavity. This increase of critical temperature depends on the
ratio between the parameter of atom-atom interaction, $\chi $, and atom
transition frequency, $\omega _{21}$. The calculation of average value of
atomic population in two different ways emphasizes the agreement between the
initial effective Hamiltonian (2) and approximation Hamiltonian (6) taken in
thermodynamical limit. Hence we conclude that the model Hamiltonian (6)
describes accurately the behavior of phase transition of quasi-spin system
with two-photon exchange interaction where a non-traditional temperature
dependence of the order parameter is manifested.

The discussed problem may be regarded as being of academic interest due to
the fact that such temperature dependence can be realized in more
complicated non-linear exchanges between the spins through the thermostat.
On the basis of two-quantum exchange mechanism between the quasi-spins, in
this paper, we pay attention to the possibility of increasing the
correlation with temperature in the system.

\bigskip

{\LARGE Figure Captions}

\bigskip

Fig.1 \ The\ temperature dependence of order parameter for different values
of ratio $\chi /\omega _{21}$: a) $\chi /\omega _{21}=0.5$; b) $\chi /\omega
_{21}=0.51$ and c) $\chi /\omega _{21}=0.6$. Here the temperature is
represented in relative units $T/T_{\mathrm{cr}}^{\max }$, where $T_{\mathrm{%
cr}}^{\max }$ corresponds to the critical temperature in phase transition of
proposed model. The dotted line corresponds to traditional phase transition
of a quasi-spin system with constant exchange integral, $\lambda $ (for
example Dicke model) and full line evidences the phase transition in the
case of the proposed model.

\bigskip

Fig.2 \ The temperature dependencies of the average value of atomic
population calculated using Exp.(4) - dotted line, and by Exp.(10) -
solid line, for case $\chi /\omega _{21}=0.6$.

\end{document}